\documentclass[12pt,a4paper]{article}
\usepackage{amssymb,amsmath,graphicx,epsfig,subcaption,cite}
\usepackage{epstopdf,xcolor}
\usepackage[utf8]{inputenc}
\usepackage[utf8]{inputenc}

\begin{document}

\begin{center}{\Large {\bf Quantum Ising Heat Engines: A mean field study}}\footnote{Dedicated to the loving memory of Prof. Amit Dutta, Indian Institute of Technology, Kanpur}

\vskip 0.8cm

{Muktish Acharyya$^{a,*}$ and Bikas K. Chakrabarti$^{b,\dagger}$}\\

\vskip 0.2cm

{\it $^{a}$Department of Physics, Presidency University, Kolkata-700073, India}\\

\vskip 0.1cm

{\it $^{b}$Saha Institute of Nuclear Physics, Kolkata 700064, India}\\

\vskip 0.1cm

$^{*}$muktish.physics@presiuniv.ac.in\\
$^{\dagger}$bikask.chakrabarti@saha.ac.in\end{center}
 
\vskip 2cm

\noindent {\bf Abstract:} We have studied the efficiencies of both classical and quantum heat
engines using an Ising model as working fluid and the mean field equation
for its non-equilibrium dynamics, formulated earlier\cite{acs,ac} to study
the dynamical hysteresis and the dynamical phase transitions in the quantum Ising ferromagnets. We studied numerically the Ising magnet's nonintegrable coupled nonlinear first order differential equations of motion for
a four stroke heat engine and compared the
efficiencies in both classical and quantum
limits using the quasi-static approximation. In both the pure classical and pure quantum cases,
the numerically  calculated  efficiencies are much
less than the corresponding Carnot values. Our
analytical formulations of the  efficiencies (both
in pure classical as well as in pure quantum Ising
heat engines) are found to agree well with the
numerical estimates. Such formulations also indicate
increased efficiency for the mixed case of a
transverse field driven Ising  engine in presence of
nonzero longitudinal field. We also numerically checked and found that
the efficiency of such a (mixed) quantum Ising
heat engine can indeed have much higher
efficiency for appropriate values of the
transverse and longitudinal fields. 
\vskip 2cm


\newpage

\noindent {\bf I. Introduction}

\vskip 0.1cm

\noindent Starting from the earliest proposal for
a quantum heat engine \cite{scovil} (using a three
level Maser) and the first proposal for
the quantum version of thermodynamics,
defining heat and work in the quantum
regime \cite{alicki1}, successive developments
\cite{feldmann1,feldmann2,alicki3,alicki4}, and also the timely reviews
\cite{rev1,rev2,rev3,rev4} led
to the recent resurgence  of theoretical model studies for quantum heat engines and
their efficiencies in comparison with almost two-century old and
thoroughly established classical heat engines which have since become a
part of our every-day life.  These
engines typically consist of a ‘heat reservoir’ (at high temperature), a
heat sink (at lower temperature) and a ‘working fluid’,  which facilitates
the (irreversible) process of extracting work from heat. Classical heat
engines and refrigerators being the major source of mechanical energy or
work in industry as well as in household, the need for miniaturization of
such heat engines also led to the quest for quantum heat engines, where
the working fluid will be a quantum system (likely many-body). 

In this paper we consider a quantum heat
engine, where the working fluid is a quantum
Ising magnet (more specifically, Ising model
in transverse field; see e.g.,\cite{bkc1,bkc2}), for
which the dynamics of Ising spins (average
magnetization) follow a mean field equation,
developed in connection with the study of
quantum hysteresis\cite{acs,ac}. These dynamical
equations for similar open quantum systems
have been compared with other formulations
(see e.g.,\cite{sarkar} and references therein) and
also discussed in connection with dynamic
simulation of materials for near-term quantum
computers (see e.g., \cite{oftelie1,oftelie2}). It may be
mentioned that some analytical results for
a quantum heat engine, using a one
dimensional Transverse Ising system (as
working fluid), has already been reported
\cite{piccitto} and the work efficiency was seen to
increase near the critical point. However,
their results are extremely limited due to
the one dimensional constraint of the spin
system considered. Our working fluid system
has in general higher dimensional quantum
quantum many-body (non-integrable) feature,
though studied here using mean field dynamics
for the engine strokes, which is not Otto
type. In the nonintegrable system considered here, we are essentially
solving the coupled nonlinear differential equation \cite{acs,ac}. Apart
from changing the temperature, we have two distinct strokes (first
and the third strokes), where the field is changing. In these strokes,
the temperature (a parameter in the equations of magnetisation
dynamics) is kept fixed. These are basically the ``isothermal"
strokes of our quantum heat engine and the
working fluid system (here quantum Ising
magnet) is allowed to exchange heats with
the source or sink. The other two strokes
where temperature changes are assumed to
be ``adiabatic". Our numerical study here
does not indicate any significant
dependence of the engine efficiency with
the ratio of durations of these ``adiabatic"
and ``isothermal" strokes. Though the
efficiency of pure quantum Ising engines
are seen to be quite small, the engine
efficiency of such quantum Ising engine
can approach the Carnot value for optimal
presence of longitudinal field. This is
consistent with the observation\cite{piccitto} in one
dimensional quantum Ising heat engine. We
believe, though a mean field study, this many-body quantum heat engines with higher
dimensional working fluid systems can be
useful for wider applicability and
comparisons, as in many nointegrable
condensed matter dynamical systems studied
in different contexts.

\vskip 1cm

\noindent {\bf II. Mean field equation and numerical solution}

\vskip 0.1cm

\noindent The Hamiltonian of Ising ferromagnet in presence of both transverse and
longitudinal external magnetic field is represented as

\begin{equation}
\mathcal {H} = -\sum_{ij} J_{ij} \sigma_i^z \sigma_j^z-h\sum_i \sigma_i^z -\gamma\sum_i \sigma_i^x.
\label{hamiltonian}
\end{equation}

\noindent Here, $\vec \sigma$ denotes the Pauli spin matrix, $h$ and $\gamma$ are the external longitunial and transverse fields respectively,
$J_{ij}$ is the ferromagnetic interaction strength between the spins placed at $i$-th and $j$-th sites. It may be noted here that due to the presence of transverse field (noncommuting component of the Hamiltonian with $\sigma_z$), the quantum nontrivial dynamics of $\sigma^z$ arises from
standard Heisenberg equation of motion. However, one can expect a simplified form of the dynamical evolution in the 
mean field approximation\cite{acs,ac}.

The mean field Hamiltonian can be written as

\begin{equation}
\mathcal {H}  \cong \sum {\vec h_{eff}} \cdot {\vec \sigma},
\label{energy}
\end{equation}

\noindent with the effective magnetic field 

\begin{equation}
{\vec  h_{eff}} \cong (m_z +h){\hat z} + \gamma {\hat x},
\label{field}
\end{equation}

\noindent where $m_z = ~<\sigma^z>$, the $z$-component of average magnetisation. The magnitude of the effective magnetic field is

\begin{equation}
|{\vec h_{eff}}| \cong \sqrt{(m_z+h)^2+\gamma^2}.
\label{modfield}
\end{equation}

\noindent The generalized mean field dynamics of the
Ising ferromagnet in presence of both
longitudinal and transverse field, extending
the classical Suzuki-Kubo formalism
\cite{suzuki}, can be represented\cite{acs} by the
following  differential equation:

\begin{equation}
\tau{{d{\vec m}} \over {dt}} = -{\vec m} + {\rm tanh}\left({{|h_{eff}|} \over {T}}\right){{\vec h} \over {|h_{eff}|}}.
\label{eqn-vecm}
\end{equation}

\noindent The above vector differential equation is basically first order nonlinear 
coupled differential equations of $m_x (=~<\sigma^x>$) and $m_z (=~<\sigma^z>$). They can be written as

\begin{equation}
\tau {{dm_x} \over {dt}} = -m_x + {\rm tanh}\left({{|h_{eff}|} \over {T}}\right)
{{\gamma} \over {|h_{eff}|}}
\label{eqn-mx}
\end{equation}

\noindent and

\begin{equation}
\tau {{dm_z} \over {dt}} = -m_z + {\rm tanh}\left({{|h_{eff}|} \over {T}}\right)
{{(m_z+h)} \over {|h_{eff}|}}.
\label{eqn-mz}
\end{equation}

For the classical case ($\gamma = 0$), in
the static limit ($dm_z/dt = 0$), eqn. (\ref{eqn-mz})
gives the Bragg-William form
of the mean field Ising equation of state
(see e. g., \cite{chaikin, stanley}), while in the dynamic case
extensive studies (see e.g.,\cite{rmp,redner}) of
dynamic hysteresis and other phenomena in
classical Ising model have been studied.
For the quantum case (with $h = 0$) the
static limit ($dm_x/dt = 0$) of eqn. (\ref{eqn-mx})
had been utilized extensively to model the
order-disorder transition behavior in
ferro-electric materials\cite{gennes} (see also
\cite{bkc1}).

These two coupled first order nonlinear differential equations have been
solved by fourth order Runge-Kutta method\cite{scarborough} with time step 
$dt = 10^{- 2}$, so that the order of the local error is significantly small.
 The initial conditions are $m_z(t = 0) = 1.0$ and
$m_x(t = 0) = 0.0$. We have calculated the instantaneous components of the
magnetization, i.e., $m_x(t)$ and $m_z(t)$. The instantaneous internal energy
($U = m_z^2 + h m_z + \gamma m_x$) of the
system has also been calculated using eqns. (\ref{eqn-mx}) and (\ref{eqn-mz}) for different drives.

\vskip 1cm

\noindent {\bf III. Four stroke Classical and quantum Ising heat engines}

\vskip 0.2cm

\begin{figure}[h]
\begin{center}

\resizebox{9cm}{!}{\includegraphics[angle=0]{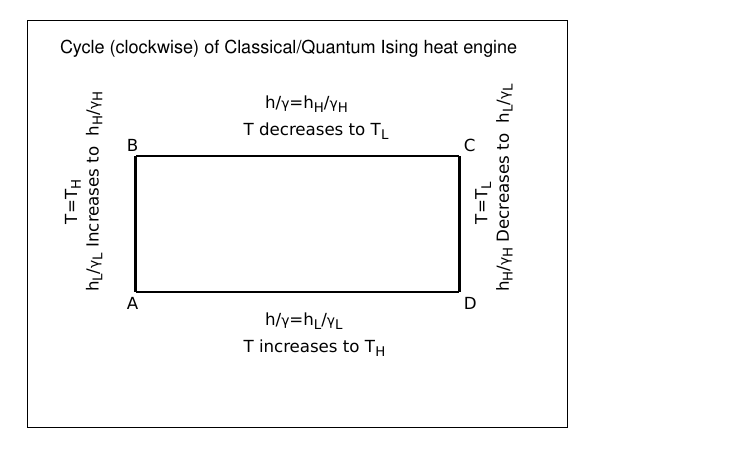}}
\caption{A schematic diagram of the cycle of engine. This starts from A and returns to A after a clockwise rotation. A schematic diagram of the for strokes AB, BC, CD and DA of the engines.
For classical Ising heat engine, stroke AB corresponds to fixed high
temperature $T = T_H$ while the longitudinal field  $h$ changes from $h_L$
to $h_H$, stroke BC corresponds to fixed high field $h = h_H$ while the
temperature $T$ changes from $T_H$ to $T_L$, stroke CD corresponds to
fixed low temperature $T = T_L$ while the longitudinal field  $h$ changes
from $h_H$ to $h_L$, and finally the fourth stroke DA corresponds to fixed
field $h = h_L$ while the temperature  $T$ changes from $T_L$ to $T_H$.
For the quantum Ising heat engine, the first stroke AB corresponds to
fixed high temperature $T = T_H$ while the transverse field  $\gamma$
changes from $\gamma_L$ to $\gamma_H$, stroke BC corresponds to fixed high
transverse field $\gamma = \gamma_H$ while the temperature $T$ changes
from $T_H$ to $T_L$, stroke CD corresponds to fixed low temperature $T =
T_L$ while the transverse field  $\gamma$ changes from $\gamma_H$ to
$\gamma_L$, and finally the fourth stroke DA corresponds to fixed
transverse field $\gamma = \gamma_L$ while the temperature  $T$ changes
from $T_L$ to $T_H$.}
\label{schematic}
\end{center}
\end{figure}

\noindent A schematic diagram for the four strokes of a complete cycle for both
classical and quantum heat engines are shown in Fig.~\ref{schematic}. The details of the
four strokes of any cycle of the Ising engine, both classical and quantum, are given in
the figure caption. In the following, we
present our results separately for the
two cases of this mean field Ising heat
engine, namely (a) classical (where
the transverse field $\gamma = 0$) and
(b) quantum (where the longitudinal field
$h = 0$).

\vskip 0.1cm

\vskip 0.1cm

 A complete cycle of the
engine (shown in Fig.~\ref{schematic}) consists of the
following four strokes (each having equal
duration, much higher than microscopic
relaxation time $\tau$):

\noindent (i) A$\to$B: Field (longitudinal $h$, in
classical case or transverse field
$\gamma$, in quantum case) increases
linearly with time from a low value,
$h_L$ or $\gamma_L$, to a high value,
$h_H$ or $\gamma_H$) at a constant
high temperature $T_H$ of the heat
bath. Heat is absorbed by the engine
during this stroke. This is an isothermal stroke. The system is absorbing the heat
from the heat source. The internal energy of the system increases in this
stroke. As in the Carnot engine, this absorbed heat pushes the piston outward
and the engine produces mechanical work.

\noindent (ii) B$\to$C: Thermalization with the cold
bath(heat sink) at temperature $T_L$. The
field remains fixed but the temperature of the system decreases linearly from $T_H$
to $T_L$.

\noindent (iii) C$\to$D: The field decreases linearly
from the high value ($h_H$ or $\gamma_H$) to
the low value ($h_L$ or $\gamma_L$).
Temperature remains fixed at $T_L$. The heat
is being released in this stroke. This is also an isothermal stroke. The system is releasing heat
to the heat sink. The internal energy is found to decrease in this
stroke. In the
Carnot cycle of this isothermal stroke (at lower temperature) the piston is
being pulled inward by the engine (by utilizing a part of the mechanical work
produced by the engine).

\noindent (iv) D$\to$E: The field remains fixed (at $h_L$
or $\gamma_L$) and the temperature increases
linearly from $T_L$ to $T_H$.

The system returns to its original (initial)
state after the completion of a cycle. Hence,
the thermodynamic state of the system denoted
by A and that denoted by E are same in all
respect. A schematic diagram of the cycle is
shown in Fig.~\ref{schematic}.

The change in the internal energy would
provide the heat absorbed by the system
and heat released by the system. Here, the
heat will be absorbed by the system in the
first stroke (A$\to$B). The heat absorbed
$E_{absorbed} = U(B)-U(A)$, where $U(A)$
and $U(B)$ represents the internal energy
at state-A and that of state-B respectively.
The heat is released in the third stroke
(C$\to$D) by the system.  The heat
released is $E_{released} = |U(D)-U(C)|$
(to keep it positive). So, the efficiency is
$\eta= (E_{absorbed}-E_{released})/
E_{absorbed}$.

In the numerical study and thermodynamic
estimates, we have discarded many transient
cycles (typically about 49 such  cycles) and
then calculated the thermodynamic quantities
in  a stable cycle (typically on 50th cycle).
Each cycle consists of four strokes and
each stroke consists of $10000×dt$
elapsed time of the differential equation.
The maximum value of the relaxation time
$\tau$ used here is 0.2 (much less than
$10000×dt = 100$ here) to maintain the
quasi static limit of the thermodynamic
working speed of the engines.

\vskip 1cm

\noindent {\bf IV. Numerical solutions of mean
field equations and estimates of internal energy
and engine efficiency in quasi-equlibrium
approximation}

\vskip 0.3cm

\noindent {\it (IV.1) Classical case:}

\begin{figure}[h]
\begin{center}
\resizebox{9cm}{!}{\includegraphics[angle=0]{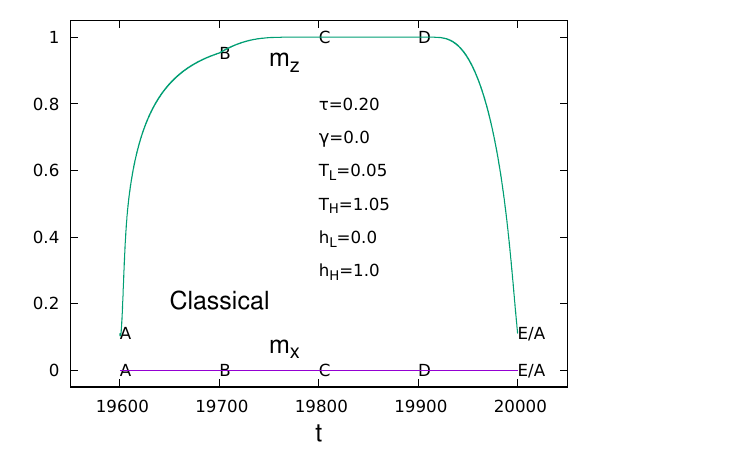}}
\caption{Numerical solutions of eqns.~(\ref{eqn-mx}) and ~(\ref{eqn-mz}) for the components ($m_x$ and
$m_z$) of the magnetization, plotted against time ($t$) over a full steady
state cycle for the classical Ising heat engine ($\gamma$ = 0).}
\label{classical-m}
\end{center}
\end{figure}

In the calculation of classical efficiency, $\eta_{C}$, we have used
$T_{L}=0.05$ and $T_{H}=1.05$ (slightly above the critical temperature
of ferro-para transition). The transverse field $\gamma=0.0$ (always). The
longitudinal field varies between $h_{L} =0$ and $h_{H}$. 
The instantaneous components of the magnetisations ($m_x$ and $m_z$) are depicted in Fig.~\ref{classical-m}. 

\begin{figure}[h]
\begin{center}
\resizebox{9cm}{!}{\includegraphics[angle=0]{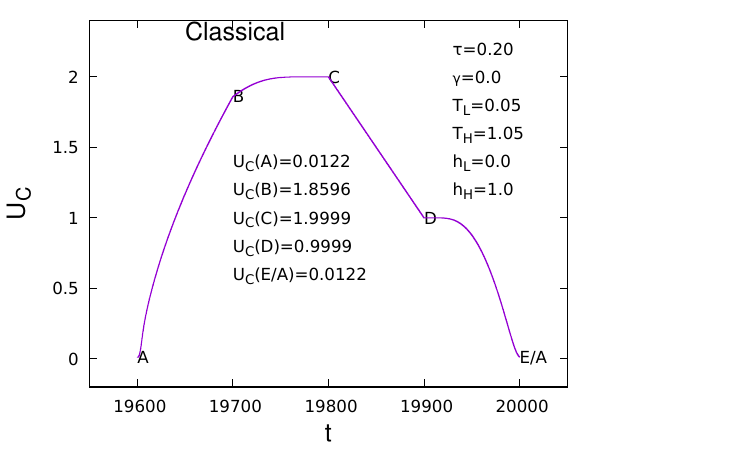}}
\caption{The numerically estimated internal energy ($U_C = m_z^2 + h m_z$) of
the working fluid (Ising magnet), plotted against time ($t$) over a full
steady state cycle for the classical Ising heat engine ($\gamma$ = 0).}
\label{classical-energy}
\end{center}
\end{figure}

In the classical calculations ($\gamma=0$) the transverse components of the magnetisation, $m_x$ will always remain zero. The dynamics will be represented by $m_z$ only.  The instantaneous internal energy ($U_C = m_z^2 +m_z h$) are shown in Fig.~\ref{classical-energy}.

\begin{figure}[h]
\begin{center}
\resizebox{9cm}{!}{\includegraphics[angle=0]{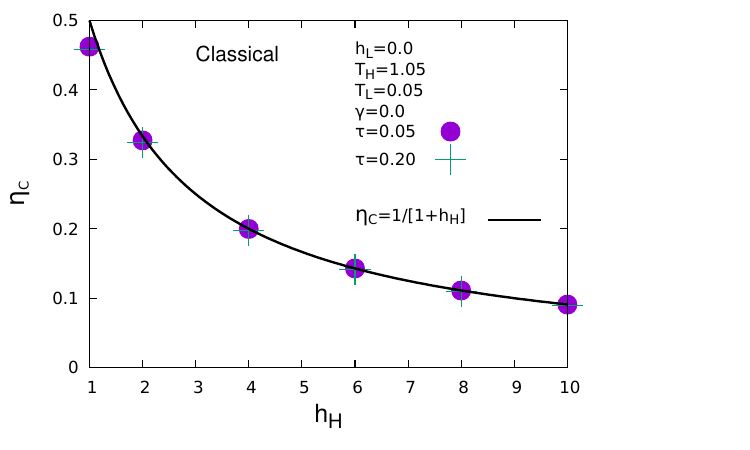}}
\caption{The efficiency $\eta_{C}$ in the case of absence of transverse field ($\gamma =0$) is plotted against the longitudinal field $h_{H}$. Two different symbols represent two different values of the microscopic relaxation time $\tau$. The solid line represents the
analytical result ($\eta_C~=~R_c/[1+h_{H}]$), where $R_c=1$.}
\label{classical-efficiency}
\end{center}
\end{figure}

We have
studied the efficiency $\eta_{C}$ as function of $h_{H}$ for two different values of the microscopic relaxation time $\tau$. The values of $\tau$ are kept small to ensure the quasi-static process is maintained in 
classical thermodynamic engines. The efficiency has been calculated from the change in the internal energy. The efficiency is studied as a function of the $h_{H}$. In our numerical simulation, we have calculated the efficiency from a steady cycle.
We have discarded 49 number of initial transient cycles. The 50th steady cycle has been used to calculate the efficiency.
Each cycle consists of four strokes. Each strokes consists of 10000 times steps (in the unit of $dt$). We have used $dt=0.01$.
 The results are shown in Fig.~\ref{classical-efficiency}. The numerical results are in good agreement with the
 theoretical prediction ($\eta_c~=~1/[1+h_H]$).

\vskip 0.3cm

\noindent {\it (IV.1a) Theoretical estimate
of classical efficiency $\eta_C$:}

\vskip 0.3cm

We estimate $\eta_C$ for the case where $T_L = 0, T_H = 1_{+}, h_L = 0$ and
$h_H > 1$.  For the classical case the energy $U_C = m_z^2 + h m_z$ and we
get (see Fig.~\ref{schematic}), using equilibrium results for the above parameter
values, $U_C(A) = 0$, $U_C(B) = 1 + h_H$, $U_C(C) = 1 +h_H$ and $U_C(D) =
1$. This gives $U_C(B) - U_C (A) = 1 + h_H$ for the heat energy taken and
$U _C(C) - U_C(D) = 1$ for the heat energy released by the classical Ising
engine, giving

\begin{equation}
\eta_C={{[(U_C(B)-U_C(A))-(U_C(C)-U_C(D)]} \over {[U_C(B)-U_C(A)]}} =
{{R_C} \over {[1+h_H]}},
\label{theoeffc}
\end{equation}

\noindent with an adjustable parameter $R_C$ of order $m_z^2$ for the dynamic
case considered in the simulations.

In Fig.~\ref{classical-efficiency} we plot the numerical estimates of the classical Ising heat
engine efficiency (for two values of the microscopic relaxation time $\tau$
in eqn.~(\ref{eqn-vecm}); here of course for numerical stability we kept $T_L = 0.05$.
The results agree fairly well with the above-mentioned theoretical
estimate of $\eta_C$ with $R_C = 1.0$.

\vskip 0.3cm

\noindent {\it (IV.1b) Classical efficiency for
nonuniform stroke distribution}

\noindent In order to check if shorter ``adiabatic” stroke
duration ($L_{Adia}$ for strokes B$\to$C and
D$\to$A in Fig. 1) compared to the duration
($L_{Iso}$) of ``Isothermal” strokes (A$\to$B and
C$\to$D in Fig. 1)) affect the classical
Ising engine efficiency, we studied numerically
for two cases ($L_{Iso} = 2L_{Adia}$ and
$L_{Iso} = 4L_{Adia}$), shown in Fig. 5.
As may be seen, we could not detect any significant
change.

\begin{figure}[h]
\begin{center}
\resizebox{9cm}{!}{\includegraphics[angle=0]{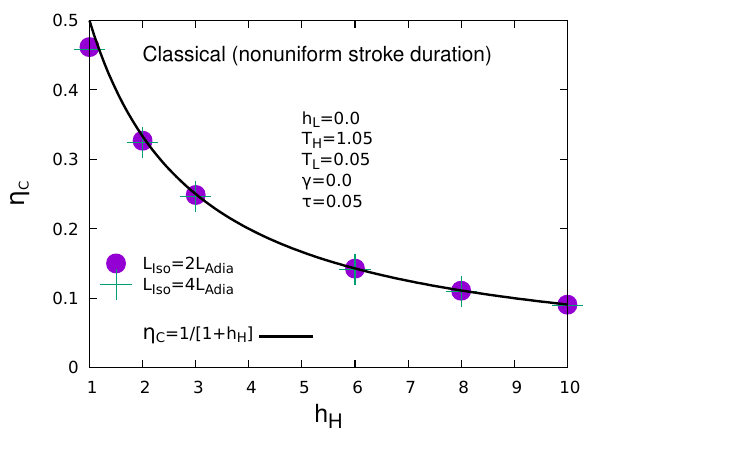}}
\caption {Classical efficiency $\eta_C$,
plotted against the longitudinal field $h_H$ for
two different fractions of the ``adibatic" stroke
duration $\L_{Adia}/L_{Iso}$, compared to the
``isothermal" stroke duration (represented by two
different symbols). The solid line again represents
the analytical result ($\eta_C = R_c/[1 + h_H])$,
with $R_c = 1$, derived above.}
\label{cl-eff-nonuniform-strokes}
\end{center}
\end{figure}

\vskip 0.2cm
\noindent {\it (IV.2) Quantum case:}

\begin{figure}[h]
\begin{center}
\resizebox{9cm}{!}{\includegraphics[angle=0]{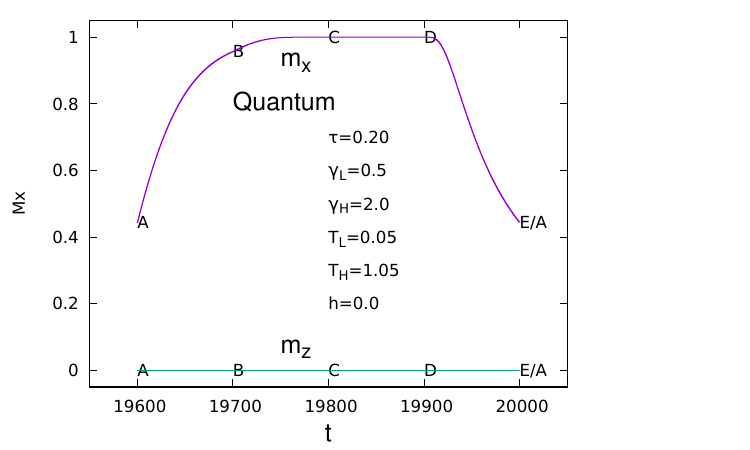}}
\caption{Numerical solutions of eqns. (6) and (7) for the components ($m_x$ and
$m_z$) of the magnetization, plotted against time ($t$) over a full steady
state cycle for the quantum Ising  heat engine (in presence of transverse
field $\gamma$ and in absence of the longitudinal field; $h$ = 0).}
\label{quantum-m}
\end{center}
\end{figure}

\noindent Our study has been extended in the presence of transverse magnetic field $\gamma$ in the absence of any strong longitudinal field
(using $h=0_{+}$) to investigate the quantum (transverse Ising ferromagnet) effects in the mean field approximation. In this case, the
transverse magnetisation ($m_x \neq 0$) plays the crucial role in determining the efficiency. Here also, like the classical case, each cycle (four-stroke) consists of 40000 time units (in the unit of $dt=0.01$).  Initial 49 number of transient cycles are discarded to achieve steady cycle. The quantities
are calculated from 50th steady cycle. The ranges of the temperatures are set as used in the case of classical study.
The transverse field operates within range $\gamma_{L}=0.5$ and $\gamma_{H}=10$. We have studied the efficiency
$\eta_{Q}$ as function of $\gamma_{H}$ keeping $\gamma_{L}=0.5$ fixed. The time dependence of the components of
the magnetizations ($m_x$ and $m_z$) are
shown in Fig.~\ref{quantum-m}. It may be mentioned here
that, accurately speaking, the longitudinal
magnetization $m_z$ has, for $h = 0_+$, a
little growth with the drive of $\gamma_H$
up to the dynamic critical point value near
unity, missed because of some subtle
instabilities.

\begin{figure}[h]
\begin{center}
\resizebox{9cm}{!}{\includegraphics[angle=0]{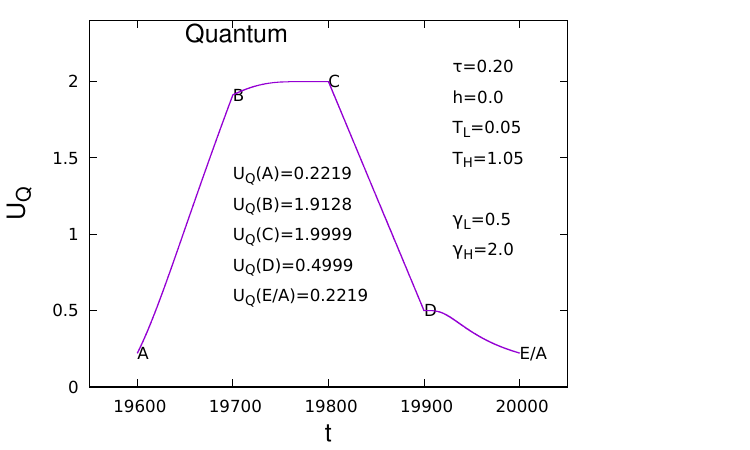}}
\caption{Numerically estimated internal energy ($U_Q = m_z^2 + \gamma m_x$) of the
working fluid (quantum Ising magnet), plotted against time ($t$) over a
full steady state cycle (for longitudinal field $h = 0$, and in the
presence of transverse field $\gamma$).}
\label{quantum-energy}
\end{center}
\end{figure}

\begin{figure}[h]
\begin{center}
\resizebox{9cm}{!}{\includegraphics[angle=0]{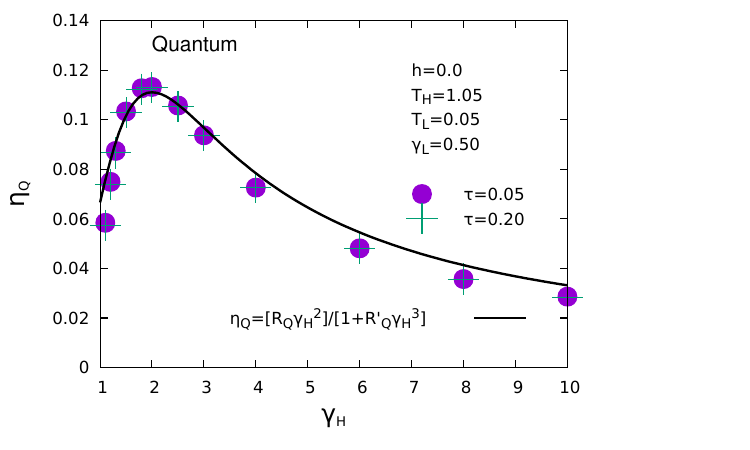}}
\caption{The efficiency $\eta_Q$ of the quantum Ising
heat engine is plotted against the peak
value ($\gamma_H$) of the transverse field
or quantum tunneling probability in the
Hamiltonian (\ref{hamiltonian}). Here, $h = 0_{+}$ and
$\gamma_L = 0.5$. Two different symbols
represent two different values of the
microscopic relaxation time $\tau$. The
solid line represents the best fit to eqn.~(\ref{theoeffq}): $\eta_Q = [R_Q\gamma_H^2]/[1 + R^{\prime}_Q\gamma_H^3]$, where $R_Q$ = 1/12 and
$R^{\prime}_Q$ = 1/4. Note that $\gamma_H = 1$ corresponds to the (equilibrium) quantum critical point (zero temperature) of the mean field system.}
\label{quantum-efficiency}
\end{center}
\end{figure}

The internal energy of the system $U_Q$ has been calculated using
the mean field approximation $U_Q \cong m_z (m_z +h) + m_x \gamma$ assuming a tiny longitudinal field
$h=0_{+}$. The instantaneous energy $U_Q$ has been shown in
Fig.~\ref{quantum-energy}. Finally, the efficiency, $\eta_{Q}$ has been studied and shown in Fig.~\ref{quantum-efficiency} as
function of $\gamma_{H}$. Unlike the classical case, here, the efficiency shows a nonmonotonic variation. The data
are fitted (see eqn.~(\ref{theoeffq})) with a function $\eta_{Q}~=~[R_{Q}\gamma_{High}^2]/[1+R'_{Q}\gamma_{high}^3)]$, where $R_{Q}=1/12$ and $R'{Q}=1/4$.
This shows good agreement with analytical prediction.

\vskip 0.2cm

\vskip 0.2cm

\vskip 0.2cm

\vskip 0.5cm

\noindent {\it (IV.2a) Theoretical estimate
of efficiency $\eta_Q$:}

\vskip 0.3cm

We estimate $\eta_Q$ theoretically for the case where
$T_L = 0, T_H = 1_{+}, \gamma_L = 0$ and $\gamma_H >> 1$.
For this quantum case, in order to see the competition
between the classical or longitudinal order ($m_z$)
explicitly  with the quantum or transverse order ($m_x$)
of the Ising manetization, we consider a tiny longitudinal field
($h = 0_{+}$) for calculating the internal energy
$U_Q = m_z^2 + \gamma m_x$. We then get (see Fig. 1), using the
quasi-equilibrium results for the above
parameter values, $U_Q(A) = 0$, $U_Q(B)
= U_Q(C) \simeq 1/\gamma_H^2 + \gamma_H$
(as $m_z^2 \simeq 1/\gamma_H^2$ from eqn.
(\ref{eqn-mz}), $m_x \simeq 1 - m_z^2/2\gamma_H^2$
from eqn. (\ref{eqn-mx}), for $\gamma_H >> 1$ and
the constraint $m_x^2 + m_z^2 =
{\rm tanh}^2(|h_{eff}|/T) \simeq 1)$, and
$U_C(D)$ = 1

\begin{equation}
\eta_Q={{[(U_Q(B)-U_Q(A))-(U_Q(C)-U_Q(D))]} \over {[(U_Q(B)-U_Q(A)]}}={{R_{Q}} \over {[{{1} \over {\gamma_H^2}} + R'_{Q} \gamma_H]}},
\label{theoeffq}
\end{equation}

\noindent with two adjustable parameters $R_{Q}, R'_{Q}$ both less than unity as they
are of order $m_z^2$ for the dynamic case considered in the simulations.

In Fig.~\ref{quantum-efficiency} we plot the numerical estimates of the quantum Ising heat
engine efficiency (again for two values of the microscopic relaxation time
$\tau$ in eqn.~(\ref{eqn-vecm}); here also for numerical stability we kept $T_L$ =
0.05). It agrees fairly well with the theoretical estimate (eqn.~(\ref{theoeffq})) with
the phenomenological fit values of the parameters $R_Q$ = 1/12 and $R'_Q$
= 1/4.

\vskip 0.6cm

\noindent {\it (IV.2b) Quantum efficiency for
nonuniform stroke distribution}

Again to check if shorter ``adiabatic” stroke
duration ($L_{Adia}$ for strokes B$\to$C and D$\to$A
in Fig. 1) compared to the duration ($L_{Iso}$) of
``Isothermal” strokes (A$\to$B and C$\to$D in Fig. 1)
affect the quantum Ising engine efficiency, we
studied numerically two cases ($L_{Iso}$ = 2$L_{Adia}$
and $L_{Iso}$ = 4$L_{Adia}$), shown in Fig. 9.  In
this case also, we cound not detect any
significant change of the quantum efficiency.

\begin{figure}[h]
\begin{center}
\resizebox{9cm}{!}{\includegraphics[angle=0]{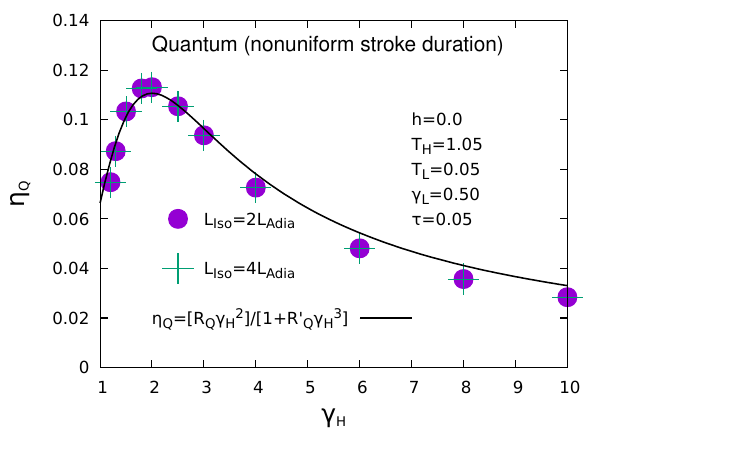}}
\caption {Quantum Ising efficiency $\eta_Q$,
plotted against transverse field $\gamma_H$ for
two different fractions of the ``adibatic" stroke
duration $\L_{Adia}/L_{Iso}$, compared to the
``isothermal" stroke duration (represented by two
different symbols). The solid line again represents
the analytical result (eqn. (9) with identical
parameter values as in Fig. 8.}
\label{eff-qm-nonuniform-strokes}
\end{center}
\end{figure}

\vskip 0.2cm

\noindent {\it (IV.2c) Quantum case with
nonzero longitudinal field:}

We have seen very low values of pure quantum
Ising engine efficiency, discussed above and
shown in Fig. 8. The theoretical estimate of
efficiency $\eta_Q$, discussed in section
(IV.2b) above indicates that the quantum
efficiency can be significantly increased in
presence of nonzero longitudinal magnetic
field ($h$). This was also suggested by the
observation\cite{piccitto} of increased efficiency in
one dimensional quantum Ising heat engine at
zero temperature. In
Fig. 10, we show the numerically estimated
efficiency $\eta_Q$ (as function of $\gamma_H$,
of the engine with the same parameters as in
Fig. 7, with $h =$ 0.1, 0.5 and 1.5. Fig. 11
shows the variations of $\eta_Q$ at transverse
field terminal value $\gamma_H$ = 1.1 and 4.0,
against the longitudinal field $h$ from Fig. 10.

The observed good fit of these data to eqn.
(9), with $\gamma_H$ replaced by $h$,  again
confirms its validity with the values of
$R_Q (\gamma_H)$ as well as of $R'_Q (\gamma_H)$.
It also indicated that the maximum quantum heat
engine efficiency $\eta_Q \simeq$ 0.78 at $\gamma_H \simeq 1.10$
with $h \simeq 0.15$). Whereas the limiting Carnot
value ($1 - T_L/T_H$) is around 0.95.

\vskip 0.5cm

\begin{figure}[h]
\begin{center}
\resizebox{9cm}{!}{\includegraphics[angle=0]{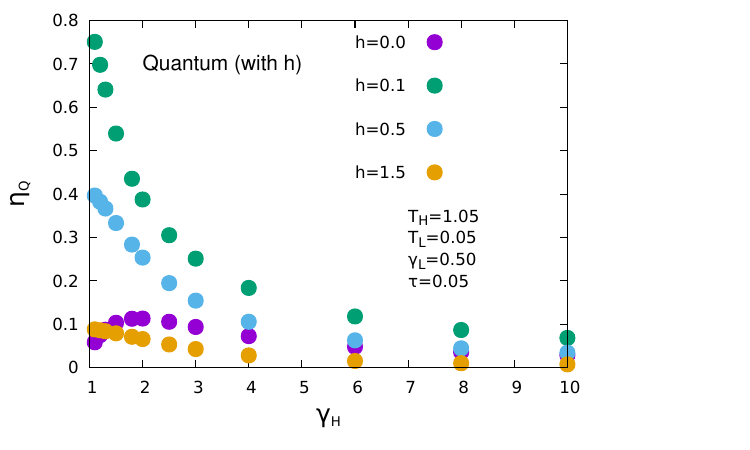}}
\caption {Quantum Ising engine
efficiency  $\eta_Q$ against the transverse field
value $\gamma_H$ for different longitudinal
field ($h$) values.}
\label{quantum-eff-gamma-diff-H}
\end{center}
\end{figure}

\vskip 0.3cm

\begin{figure}[h]
\begin{center}
\resizebox{9cm}{!}{\includegraphics[angle=0]{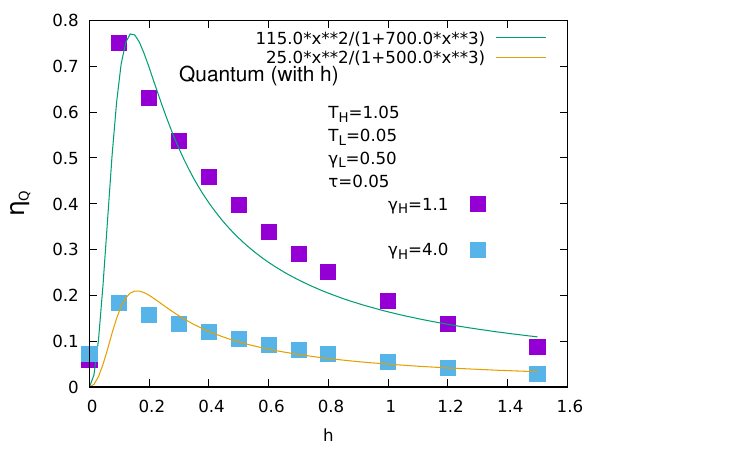}}
\caption {Plot of quantum Ising efficiency
$\eta_Q$ vs. longitudinal field strength ($h$) for
two chosen values of transverse field $\gamma_H$.
The observed fit of these data to eqn. (9), with
$\gamma_H$ replaced by $h$, indicates that the
maximum value of efficiency of such quantum Ising heat engine 
(having $T_H$ = 1.05 and $T_L$ = 0.05) 
becomes $\eta_Q \simeq 0.78$ at $\gamma_H \simeq 1.10$  with $h \simeq 0.15$.
Whereas, the Carnot efficiency ($1 - T_L/T_H$) is approximtely equal to 0.95.}
\label{quantum-eff-H-diff-gamma}
\end{center}
\end{figure}

\newpage

\noindent {\bf V. Summary and discussions}

\vskip 0.2cm

\noindent We studied here numerically (using Runge-Kutta method) the generalized
mean field dynamical equations of the longitudinal ($m_z$) and transverse
($m_x$) components of the magnetization, using a generalized form (\cite{acs},
see e.g., \cite{ac,bkc1} for details) of the  Suzuki-Kubo formalism\cite{suzuki} for the
classical Ising magnet, for a four stroke (open and irreversible) quantum
Ising heat engine (see Fig.\ref{schematic}). We also estimated, using quasi-equilibrium
approximation, the efficiencies of the heat
engines in {both classical and quantum limits (see eqns. (8)
and (9)). We find that their generic forms agree
well (see Figs. 4 and 8) with our numerical
estimates. We tried to check if shorter ``adiabatic"
stroke duration ($L_{Adia}$ for strokes B$\to$C and
D$\to$A in Fig. 1), compared to the duration
($L_{Iso}$) of ``Isothermal" strokes (A$\to$B
and C$\to$D in  Fig. 1), would increase the
classical or quantum Ising engine efficiencies.
Our numerical studies, for two cases
($L_{Iso} =2L_{Adia}$ and  $L_{Iso} =  4L_{Adia}$),
shown in Fig. 5 (classical case) and Fig.9 (quantum
case) did not indicate any significant change of
quantum Ising engine efficiencies. In both cases, the
efficiencies are much less than the corresponding
Carnot values ($ {\eta}_{Carnot} = 1 - {T_L/T_H }$) for
temperatures $T_H = 1.05$ of the heat reservoir
and $T_L = 0.05$ of the heat sink in the classical
and quantum engines considered here. For quantum Ising heat engine, however, the
efficiency $\eta_Q$ has been found to increase significantly (see section IV.2c) in presence of optimal value
of the longitudinal field.
This is also consistent with the observation
\cite{piccitto} of increased efficiency in presence of
optimal value of $h$ in one dimensional ($T$
= 0) quantum Ising engine. As may be seen from
Fig. 11, the best fit the analytical form (9)
 of the quantum Ising heat
engine efficiency (with $\gamma_H$ replaced by
$h$ in the equation and using $\gamma_H$ dependent
fitting parameters  $R_Q$ and $R'_Q$) becomes maximum ($\eta_Q \simeq 0.78$) at $\gamma_H \simeq 1.10$ with $h \simeq 0.15$.
We also note, similar to  the observation of a shift
from the equilibrium quantum critical point for
the maximum efficiency point in the one dimensional
quantum Ising heat engine\cite{piccitto}, the efficiency of
such mean field quantum Ising heat engine also becomes
maximum (see Fig. 8) above the equilibrium quantum
critical point.

It may be noted here, that the cycles described for such a quantum Ising
heat engine (studied by solving the coupled nonlinear differential equations)
are different from those of the Otto-like cycles for integrable one dimensional
Ising engine considered earlier \cite{piccitto}. We have calculated
the efficiency
just from the amounts of heat absorbed and released by the engine. The first
stroke (A$\to$B, isothermal and field increasing) provides the heat absorbed
from the heat reservoir by the quantum Ising system. The third stroke (C$\to$D)
(isothermal again where the field decreases) induces the heat released
to the heat
sink. The work done by the engine is basically the difference in heat absorbed and heat released. The engine efficiency has been calculated (as the ratio of work done and the heat absorbed by the engine) using the internal
energy changes
during these two strokes.

Last but not the least, the study of
such dynamical behavior of the quantum
heat engines using nonintegrable many-body
condensed matter systems\cite{rev2} (even with
mean field approximation) are expected to
be interesting and useful. The nonintegrable mean field equations
here are simple (even compared to that of the quantum Heisenberg
equation of motion for its unitary evolution). Moreover, the solutions
of these nonlinear and nonintegrable mean field equations can
provide the analysis of some instabilities at appropriate parameter
values (of temperature, longitudinal
field and transverse field) and give the
crossover indications from its (quantum)
heat engine to refrigerating engine behavior
(even perhaps though supercritical pitchfork
bifurcation\cite{strogatz}).
 It would also be interesting
to study if such bifurcations are indeed there
and can be observed. This would of course require considerably increased
accuracy in the numerical solutions of the
nonlinear dynamical equations  for the quantum phase
transition  in the presence of the
noncommuting time-dependent transverse field (see discussions
in subsection IV.b).

\vskip 0.5cm

\noindent {\bf Acknowledgements:} Amit Dutta wanted us to explore the application of
quantum mean field equation we had studied earlier in the context of dynamic
hysteresis
in quantum Ising systems to quantum Ising heat engines. He promised to teach us about the development and literature on quantum
heat engines. We missed that opportunity because of his very sudden and
untimely demise. We are thankful to Heiko Rieger and Eduardo Hernandez,
former and present Editor-in-Chief of European Physical Journal B (EPJB),
for taking the initiative of this Topical Issue on ``Quantum phase
transitions and open quantum systems: A tribute to Prof. Amit Dutta", in
memory of their one long time Editor.  We would like to thank the Guest
Editors of this Special Issue of EPJB, Uma Divakaran, Ferenc Iglói, Victor Mukherjee and Krishnendu
Sengupta for kind invitation to contribute in it. We are  extremely thankful to Victor Mukherjee for
a careful reading of the paper and giving important
suggestions. We are indebted to an anonymous
referee for suggesting an extended study of the
quantum Ising engine under external field. MA is grateful to
the Presidency University for the FRPDF grant and BKC is grateful to the
Indian National Science Academy for their Senior Scientist Research
Grant.

\vskip 0.2cm

\noindent {\bf Conflict of interest statement:} We declare that this manuscript is free from any conflict of interest. The
authors have no financial or proprietary interests in any material discussed in this article.

\vskip 0.2cm

\noindent {\bf Data availability statement:} The data may be available on request to the corresponding author.

\vskip 0.2cm

\noindent {\bf Funding statement:} No funding was received particularly to support this work.

\vskip 0.2cm

\noindent {\bf Authors’ contributions:} Bikas K. Chakrabarti-conceptualized the problem,  analysed the
results, developed the approximate analytic formulations, wrote the manuscript. Muktish Acharyya-developed the code for numerical simulation, collected the data, prepared the figures, analysed the results,
wrote the manuscript.

\vskip 0.5cm


\end{document}